\documentclass[10pt]{article}

\usepackage{geometry}
\usepackage{graphicx} % zyx
\usepackage{multirow}
\usepackage{threeparttable}
\usepackage{makecell}
\usepackage{color}
\usepackage{cite}

\title{Multimodal Affective States Recognition Based on Multiscale CNNs and Biologically Inspired Decision Fusion Model}

\author
{Yuxuan Zhao$^{1,2}$, Xinyan Cao$^{1}$, Jinlong Lin$^{1}$, Dunshan Yu$^{1}$, Xixin Cao$^{1\star}$\\
\\
\normalsize{$^{1}$School of Software and Microelectronics, Peking University.}\\
\normalsize{$^{2}$Institute of Automation, Chinese Academy of Sciences.}\\
\\
\normalsize{To whom correspondence should be addressed; E-mail:  cxx@ss.pku.edu.cn.}
}

\begin{document}

\date{}

\baselineskip15pt

\maketitle

\begin{abstract}
There has been an encouraging progress in the affective states recognition models based on the single-modality signals as electroencephalogram (EEG) signals or peripheral physiological signals in recent years. However, multimodal physiological signals-based affective states recognition methods have not been thoroughly exploited yet. Here we propose Multiscale Convolutional Neural Networks (Multiscale CNNs) and a biologically inspired decision fusion model for multimodal affective states recognition. Firstly, the raw signals are pre-processed with baseline signals. Then, the High Scale CNN and Low Scale CNN in Multiscale CNNs are utilized to predict the probability of affective states output for EEG and each peripheral physiological signal respectively. Finally, the fusion model calculates the reliability of each single-modality signals by the Euclidean distance between various class labels and the classification probability from Multiscale CNNs, and the decision is made by the more reliable modality information while other modalities information is retained. We use this model to classify four affective states from the arousal valence plane in the DEAP and AMIGOS dataset. The results show that the fusion model improves the accuracy of affective states recognition significantly compared with the result on single-modality signals, and the recognition accuracy of the fusion result achieve 98.52\% and 99.89\% in the DEAP and AMIGOS dataset respectively.
\end{abstract}

%Strat
\section{Introduction}
Affective states recognition plays a crucial role in human-machine interaction and health care. The recognition method based on physiological signals has become a research hotspot because the signals could represent the affective states and cannot be controlled subjectively compared with other signals such as facial expressions \cite{Face}, gesture or speech. The physiological signals are composed of electroencephalogram (EEG) signals and peripheral physiological signals, and the peripheral physiological signals include electrocardiogram (ECG) signals, electromyogram (EMG) signals, galvanic skin resistance (GSR) signals, etc. Most of the studies of affective states recognition based on physiological signals only focus on either EEG signals or peripheral physiological signals, and they ignore the correlation between these two signals. To solve this problem and further improve the accuracy, some researchers integrate the EEG signals and peripheral physiological signals to predict affective states. The research on affective states recognition from multimodal physiological signals can be categorized into feature level fusion \cite{RN310,RN520,RN425,RN627}, intermediate level fusion \cite{RN630,RN312}, and decision level fusion \cite{RN629,RN679,RN677}. Although these studies have been achieved, there still exist some challenging problems. Firstly, few studies have focused on the decision level fusion, and most of the decision level fusion methods are the voting methods. Secondly, other studies \cite{RN630,RN312,RN520,RN425,RN678,RN679,RN627,RN629,RN683,RN486,RN682,RN681,RN677} have failed to achieve a high recognition accuracy which is not enough for implementing real-world applications.

To address these issues, in this paper, we propose a Multiscale Convolutional Neural Networks (Multiscale CNNs) and a biologically inspired decision fusion model for affective states recognition based on multimodal physiological signals. The High Scale CNN and Low Scale CNN in the Multiscale CNNs are utilized to EEG- and peripheral physiological-based affective states recognition respectively.

The main contributions of the proposed work are summarized as follows: 

(1) The proposed Multiscale CNNs could predict affective states effectively from various physiological signals. 

(2) The proposed work introduces a simple and effective decision level fusion method which improves the accuracy of affective states recognition significantly compared with the result on single-modality signals.

(3) The proposed fusion model is inspired by the multisensory integration studies from neuroscience and psychophysiology, and the decision is made by the more reliable modality information while other modalities information is retained. Therefore, it is more effective than other decision level fusion methods such as nondiscriminatory plurality voting, and it could be applied to other multimodal recognition tasks.

The remainder of this paper is organized as follows: Section 2 provides a review of research about the multimodal information fusion techniques used in multimodal physiological signals-based affective states recognition tasks and the multisensory integration studies in neuroscience and psychophysiology. Section 3 describes the method of data pre-processing, the multiscale convolutional neural networks, and the biologically inspired decision fusion model. Section 4 presents the description of the DEAP and AMIGOS dataset which are used in the experiment, the fusion performance analysis of the biologically inspired decision fusion model, and the comparison with existing models. In Section 5, we conclude our work.

\section{Related works}

\subsection{Multimodal information fusion techniques}
Multimodal information fusion techniques are used to improve the performance of the system by integrating different information at the feature level, intermediate level and decision level \cite{RN623,RN434}. 

The feature level fusion techniques are widely used in multimodal affective states recognition tasks \cite{RN310,RN520,RN425,RN627,RN678}. Verma et al. \cite{RN310} proposed a multimodal fusion approach at the feature level, they extracted 25 features from EEG and peripheral signals by Discrete Wavelet Transform, then used the Support Vector Machine (SVM) classifier for thirteen affective states classification. Kwon et al. \cite{RN520} proposed a preprocessing method of EEG signals in which they used a wavelet transform while considering time and frequency simultaneously, then extracted the EEG feature convolution neural networks (CNN) model and combined them with the features extracted from galvanic skin response (GSR) signals for affective states recognition. Hassan et al. \cite{RN425} applied unsupervised deep belief network (DBN) for depth level feature extraction from partial peripheral signals and combined them in a feature fusion vector, then used the Fine Gaussian Support Vector Machine (FGSVM) to classification. Zhou et al. \cite{RN627} used convolutional auto-encoder (CAE) to obtain the fusion features of various signals, then used a fully connected neural network classifier for affective states recognition. Ma et al. \cite{RN678} proposed a multimodal residual LSTM (MM-ResLSTM) network for affective states recognition. This network could learn the correlation between the EEG and other physiological signals by sharing the weights across the modalities in each LSTM layer. 

The intermediate level fusion \cite{RN630,RN312} can cope with the imperfect data reliably \cite{RN434}. Shu et al. \cite{RN630} proposed a fusion method to model the high-order dependencies among multiple physiological signals by Restricted Boltzmann Machine (RBM). The new feature representation is generated from the features of EEG signals and peripheral physiological signals by RBM. Then they used the Support Vector Machine (SVM) classifier for affective states recognition. Yin et al. \cite{RN312} proposed a multiple-fusion-layer based ensemble classifier of stacked auto-encoder (MESAE) for affective states recognition. They extracted 425 salient physiological features from EEG signals and peripheral physiological signals. The features are split into non-overlapped physiological feature subsets, then got the abstraction fusion features by the SAE, and used the Bayesian model to classification. 

Compared with the feature level fusion and intermediate level, the research on decision level fusion \cite{RN629,RN679,RN677} is less and simpler. Bagherzadeh et al. \cite{RN629} extracted spectral and time features from peripheral and EEG signals, and non-linear features from EEG signals, then used multiple stacked autoencoders in a parallel form (PSAE) to primary classification, the final decision about classification was performed using the majority voting method. Huang et al. \cite{RN679} proposed an Ensemble Convolutional Neural Network (ECNN) which could automatically mine the correlation between various signals, then used the plurality voting strategy for affective states recognition. Li et al. \cite{RN677} used an Attention-based Bidirectional Long Short-Term Memory Recurrent Neural Networks (LSTM-RNNs) to automatically learn the best temporal features, then used a deep neural network (DNN) to predict the probability of affective states output for each modality, finally fused these result on decision level to get the result. Compared with the feature level fusion and intermediate level, the decision level fusion is more easily because the decisions resulting from multiple modalities usually have the same form of data and every modality can utilize its best suitable classifier or model to learn its features \cite{RN623}.

\subsection{Bayes-optimal cue integration model}
The multisensory integration has been widely studied in neuroscience and psychophysiology \cite{RN639}. A Bayes-optimal cue integration model has been proposed and has proved to be successful in visual and haptic information integration \cite{RN640}, visual and proprioceptive information integration \cite{RN648}, visual and vestibular information integration \cite{RN641,RN645}, visual and auditory information integration \cite{RN646,RN647}, stereo and texture or texture and motion information integration in vision research \cite{RN642,RN644}. The cue is a terminology for neuroscience. And this terminology is defined as "Any signal or piece of information bearing on the state of some property of the environment. Examples include binocular disparity in the visual system, interaural time or level differences in audition and proprioceptive signals (for example, from muscle spindles) conveying the position of the arm in space \cite{RN639}." This model estimates the result by weighting the cues in proportion to their relative reliability which is proportional to the inverse variance. Take a spatial localization estimate by visual and auditory information integration as an example as shown in Fig.\ref{b}, this model could be described as follows:

\begin{figure}
\centering
\includegraphics[width=6cm]{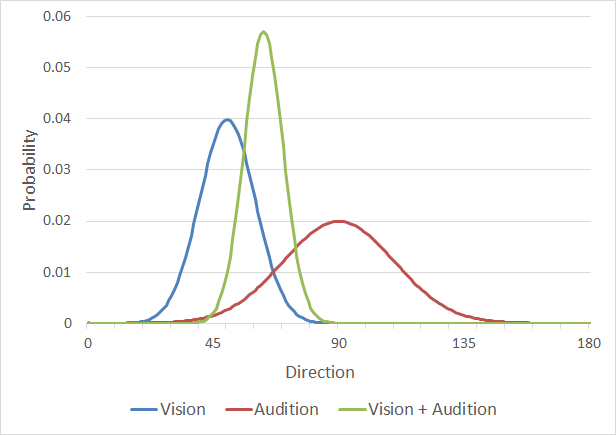}
\caption{Bayes-optimal combination of multiple sensory cues \cite{RN639,RN646}.}
\label{b}
\end{figure}

The direction of an event is s, the estimated spatial location by visual ($x_{v}$) and auditory ($x_{a}$) will inconsistent with the true location because of the noise in transmission and processing. According to Bayes' theorem and the conditionally independent sources of visual and auditory, is can be described as Equation~\ref{eqn_1}.

\begin{equation}
\label{eqn_1}
p(s|x_{v},x_{a})\propto p(x_{v}|s)*p(x_{a}|s)*p(s)
\end{equation}

If assumes that the p(s) is uniform prior distribution and the additional simplifying assumption of Gaussian likelihood function, the average estimate derived from an optimal Bayesian integrator is a weighted average of the average estimates that would be derived from each cue alone, it could be described as Equation~\ref{eqn_2}.

\begin{equation}
\label{eqn_2}
s^\ast=\omega_v\times x_v+\omega_a\times x_a
\end{equation}

where

\begin{equation}
\label{eqn_bgw}
\omega_v=\frac{\displaystyle1/\sigma_v^2}{1/\sigma_v^2+1/\sigma_a^2}\;and\;\omega_a=\frac{\displaystyle1/\sigma_a^2}{1/\sigma_v^2+1/\sigma_a^2}
\end{equation}

The sigma in Equation~\ref{eqn_bgw} is the parameter of the fitted Gaussian distribution function, and the inverse of sigma squared means the cue's reliability. In the field of neuroscience, the terminology reliability can be used as a synonym for the precision of a measurement, which is defined mathematically as its inverse variance \cite{RN639}. The greater a cue's reliability, the more it contributes to the final estimate.

\section{Methods}
\begin{figure}
\centering
\includegraphics[width=8cm]{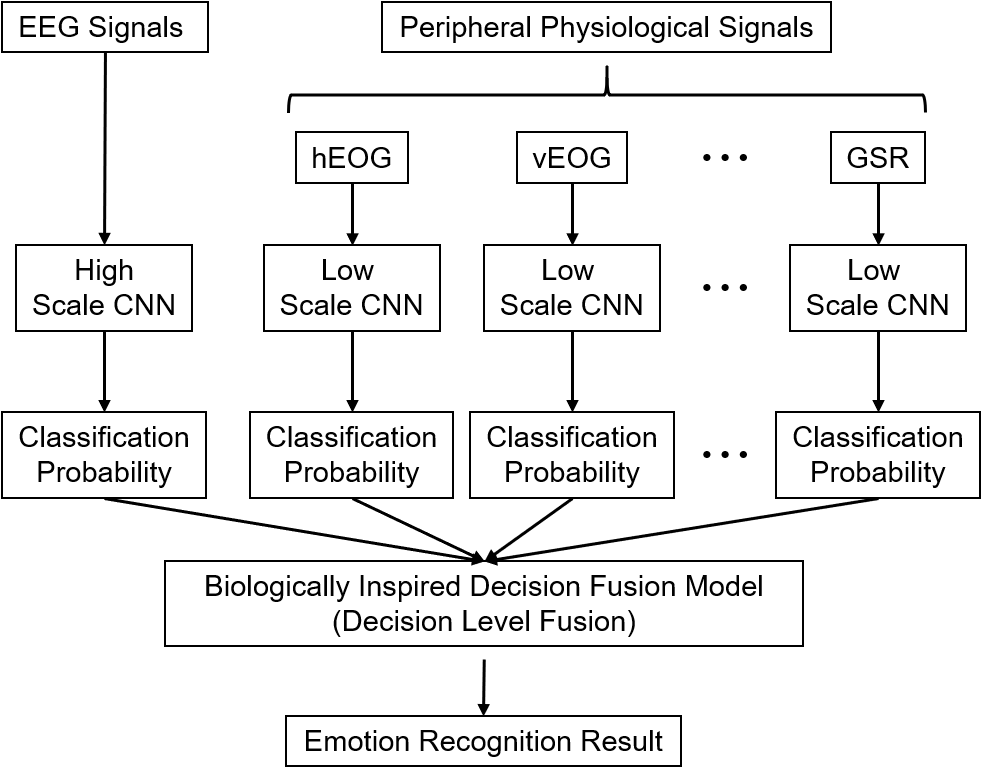}
\caption{The framework of our model.}
\label{Model}
\end{figure}

Here we propose a Multiscale Convolutional Neural Networks (Multiscale CNNs) for affective states recognition based on various physiological signals, and a biologically inspired decision fusion model that could integrate the result from Multiscale CNNs for multimodal affective states recognition. The details of the model are shown in Fig.\ref{Model}.

\subsection{Pre-processing}
A pre-processing method with baseline signals - the signals are recorded when the participant under no stimulus - which is first elaborated by Yang \cite{RN569} is an effective way to improve recognition accuracy. They reported that the pre-processing method could increase recognition accuracy by 32\% approximately in the affective states recognition task. The pre-processing method contains: extracting the baseline signals from all channels C and cut it in N segments with fixed length L, getting N segments C x L matrixes; calculating the mean value of the baseline signals with segmented data, getting the baseline signals mean value M, a C x L matrixes; cutting the EEG and Physiological signals which without baseline signals with length L and minus the baseline signals mean value M, getting the preprocessed signals.

The raw EEG signals in the dataset lost the topological position information of the electrodes. To solve this problem, the electrodes used in the dataset are relocated to the 2D electrode topological structure based on the 10-20 system positioning. For each time sample point, the EEG signals are mapped into a 9x9 matrixes. The unused electrodes are filled with zero. Z-score normalization is used in each transformation.

\subsection{Multiscale Convolutional Neural Networks}

The details of the High Scale CNN and the Low Scale CNN in the Multiscale CNNs are shown in Table \ref{MultiCNN}.

\begin{table}[]
\centering
\caption{The details of the multiscale convolutional neural network model}
\label{MultiCNN}
\renewcommand{\arraystretch}{1.5} 
\begin{tabular}{|c|c|c|c|c|}
\hline
                & \multicolumn{2}{c|}{High Scale CNN}                                                                                    & \multicolumn{2}{c|}{Low Scale CNN}                                                                                     \\ \hline
Layer Type      & \begin{tabular}[c]{@{}c@{}}Output\\ Size\end{tabular} & \begin{tabular}[c]{@{}c@{}}Kernel,Units,\\ Stride\end{tabular} & \begin{tabular}[c]{@{}c@{}}Output\\ Size\end{tabular} & \begin{tabular}[c]{@{}c@{}}Kernel,Units,\\ Stride\end{tabular} \\ \hline
Input           & 128x9x9                                               & \textbackslash{}                                               & 128x1                                                 & \textbackslash{}                                               \\ \hline
Convolutional   & 128x9x9                                               & \begin{tabular}[c]{@{}c@{}}(4x3x3), 32,\\ (1,1,1)\end{tabular} & 128x1                                                 & \begin{tabular}[c]{@{}c@{}}(3), 16,\\ (1)\end{tabular}          \\ \hline
Pooling         & 64x9x9                                                & \begin{tabular}[c]{@{}c@{}}(2x1x1),\\ (2,1,1)\end{tabular}     & 128x1                                                 & \begin{tabular}[c]{@{}c@{}}(2),\\ (1)\end{tabular}             \\ \hline
Convolutional   & 64x9x9                                                & \begin{tabular}[c]{@{}c@{}}(4x3x3), 64,\\ (1,1,1)\end{tabular}  & 128x1                                                 & \begin{tabular}[c]{@{}c@{}}(3), 32,\\ (1)\end{tabular}          \\ \hline
Pooling         & 32x9x9                                                & \begin{tabular}[c]{@{}c@{}}(2x1x1),\\ (2,1,1)\end{tabular}     & 128x1                                                 & \begin{tabular}[c]{@{}c@{}}(2),\\ (1)\end{tabular}             \\ \hline
Flatten         & 165888                                                & \textbackslash{}                                               & 4096                                                  & \textbackslash{}                                               \\ \hline
Fully Connected & 1024                                                  & \textbackslash{}                                               & 256                                                   & \textbackslash{}                                               \\ \hline
Dropout         & 1024                                                  & \textbackslash{}                                               & 256                                                   & \textbackslash{}                                               \\ \hline
Softmax         & 4                                                     & \textbackslash{}                                               & 4                                                     & \textbackslash{}                                               \\ \hline
\end{tabular}
\end{table}

The High Scale CNN which we proposed in \cite{IJCNN2020} is used for EEG-based affective states recognition. The architecture of the High Scale CNN contains two convolution layers. Each of them is followed by a max-pooling layer, and a fully-connected layer. The input size is 128x9x9, the 9x9 is the 2D electrode topological structure and the 128 is the number of the consecutive time sample point processed at once. The kernel size of the convolution layer is 4x3x3, which means the spatial-temporal features are generated based on a local topology of 3x3 and a time period of 4-time sample points.  The RELU activation function is used after the convolution operation. The pooling size of a max-pooling layer is 2x1x1 which is used to reduce the data size in the temporal dimension and improve the robustness of extracted features. To prevent missing information of input data, the zero-padding is used in each convolutional and pooling layers. The numbers of feature maps in the first and second convolutional layers are 32 and 64 respectively. Before passing the 64 resulting feature maps to the fully-connected layer, the output feature maps are reshaped in a vector. The fully-connected layer maps the feature maps into a final feature vector of 1024. And a dropout regularization after fully connected layers used to avoid overfitting. The output size is 4, which is equal to the number of labels in the task.

The Low Scale CNN is used for peripheral physiological-based affective states recognition. The architecture of the Low Scale CNN contains two convolution layers. Each of them is followed by a max-pooling layer, and a fully-connected layer. The input size is 128x1 which represents a 128 consecutive time sample point of peripheral physiological signals in each modality. The kernel size of the convolution layer is 3. The RELU activation function is used after the convolution operation. The pooling size of a max-pooling layer is 2 which is used to reduce the data size in the temporal dimension and improve the robustness of extracted features. To prevent missing information of input data, the zero-padding is used in each convolutional and pooling layers. The numbers of feature maps in the first and second convolutional layers are 16 and 32 respectively. Before passing the 32 resulting feature maps to the fully-connected layer, the output feature maps are reshaped in a vector. The fully-connected layer maps the feature maps into a final feature vector of 256. And a dropout regularization after fully connected layers used to avoid overfitting. The output size is 4, which is equal to the number of labels in the task.

\subsection{Biologically Inspired Decision Fusion Model}
%In the research of emotion models, Russell's valence-arousal scale \cite{RN649} has been widely used to quantitatively describe emotions. After watching a video that could induce the participant's emotion, the participant rated a self-assessment of arousal and valence on a scale from 1 to 9. In the emotion recognition task, the researchers often divided the dataset into various classes based on the threshold of arousal and valence. So the class labels could mapping in the two-dimensional space by calculating the mean value of the corresponding class labels on Arousal-Valence data. In this way, we can calculate the Euclidean distance between various class labels, then using the standard normal distribution to calculate the classification score reliability of other labels in a certain label based on the Euclidean distance, as shown in Equation~\ref{eqn_dg}.

In the biological example of spatial localization estimate by visual and auditory information integration, each cue evaluates the position of the object in a series of consecutive angles, and the response of sensory neurons to stimuli such as visual cue and auditory cue obey the Gaussian distribution. But in the traditional classification model, there is no relationship between different labels, and the classification scores of different labels do not obey the Gaussian distribution. Because the proposed fusion model is inspired by the Bayes-optimal cue integration model, which has been widely studied in neuroscience and psychophysiology, we ensure the biological rationality of the proposed fusion model in the following ways:

(1) Build relationships between various class labels. 
In the research of affective states models, Russell's valence-arousal scale \cite{RN649} has been widely used to quantitatively describe affective states. After watching a video that could induce the participant's affective states, the participant rated a self-assessment of arousal and valence on a scale from 1 to 9. In the affective states recognition task \cite{RN630,RN312,RN520,RN425,RN678,RN679,RN627,RN629,RN683,RN486,RN682,RN681,RN677}, the researchers often divided the dataset into various classes based on the threshold of arousal and valence. So the class labels could be mapped in the two-dimensional space by calculating the mean value of the corresponding class labels on Arousal-Valence data. We use the participants' rating scores of arousal and valence to represent the four class labels. Then we build the relationships between various class labels with the Euclidean distance of these rating scores.

(2) Calculate the classification score reliability between label $i$ and label $j$, as shown in Equation~\ref{eqn_dg}. The $f(d_{ij})$ is the classification score reliability between label $i$ and label $j$. If the label is $i$, we take the Euclidean distance between label $i$ and other labels as the variable to construct a standard normal distribution, and calculate the classification score reliability of each label when the label is $i$. For example, the dataset is divided into 4 classes based on the threshold of arousal and valence: LALV (e.g., sad), HALV (e.g., stressed), LAHV (e.g., relaxed), HAHV (e.g., happy). We can get a 4x4 classification score reliability matrix. If the current label is HAHV, according to the standard normal distribution, the classification score reliability is the highest when the classification probability is HAHV, and the lowest when the classification probability is LALV because the distance between label HAHV and label LALV is the farthest.

\begin{equation}
\label{eqn_dg}
f(d_{ij})=\frac{1}{\sqrt{2\pi}}e^{-\frac{d_{ij}^{2}}{2}}
\end{equation}

(3) In each modality,  the final classification score of each label is calculated based on the classification probability from the CNN model and classification score reliability, as shown in Equation~\ref{eqn_gr}. The $GauPR_{m,j}$ is the final classification score of label $j$ in modality $m$, the $NL$ is the number of labels, and the $PR_{m,j}$ represents the classification probability of label $j$ in modality $m$. Take the final classification score of label HAHV as an example. The final classification score of HAHV is the sum of the product of classification probability of HAHV that output by the CNN model and the classification score reliability that between label HAHV and label LALV, HALV, LAHV and HAHV.

\begin{equation}
\label{eqn_gr}
GauPR_{m,j}=\sum_{i=1}^{NL}PR_{m,j}\times f(d_{ij})
\end{equation}

%The $f(d_{ij})$ is the classification score reliability between label $i$ and label $j$. For example, the emotion dataset is divided into 4 classes based on the threshold of arousal and valence: LALV (e.g., sad), HALV (e.g., stressed), LAHV (e.g., relaxed), HAHV (e.g., happy). For a certain label HAHV, the distance between HAHV and LALV is the farthest, so the classification score reliability of LALV is lowest. Similarly, the classification score reliability of HAHV is lowest for a certain label LALV.

%In each modality, we calculate the final classification score through classification probability from the CNN model and classification score reliability as shown in Equation~\ref{eqn_gr}. The $NL$ is the number of labels, and the $PR_{c,i}$ represents the classification probability of label $i$ in modality $c$, the $GauPR_{c,j}$ is the final classification score of label $j$ in modality $c$ with classification score reliability. 

Considering the difference between the biological system and the machine systems, here we use the dispersion of final classification score to represent the corresponding modality reliability. The dispersion calculated by the standard deviation as shown in Equation~\ref{eqn_s}, the $S_{m}$ represents the modality reliability of modality $m$. The more average the final classification scores are, the smaller of the dispersion is, the lower the corresponding modality reliability is. If one of the final classification scores in a modality is high, the corresponding modality reliability is high because of the greater dispersion.

\begin{equation}
\label{eqn_s}
S_{m}=\sqrt{\frac{1}{NL-1}\sum_{j=1}^{NL}(GauPR_{m,j}-\overline{GauPR_{m}})^{2}}
\end{equation}

where

\begin{equation}
\label{eqn_x}
\overline{GauPR_{m}}=\frac{1}{NL}\sum_{j=1}^{NL}GauPR_{m,j}
\end{equation}

Then we calculate the fusion result of label $j$ as shown in Equation~\ref{eqn_f}. The $NM$ represents the number of the modalities. Then we use $argmax$ to select the final result. The argmax is a function that returns the value of variable $j$ when $F_{j}$ reaches the maximum value. For example, $F_{j}$ reaches the maximum value when the label $j$ is HAHV, and the value of variable $j$ returned by argmax is HAHV, that is, Rlabel = HAHV.

\begin{equation}
\label{eqn_f}
F_{j}=\sum_{m=1}^{NM}GauPR_{c,j}\times S_{m}
\end{equation}

\begin{equation}
\label{eqn_r}
R_{label}=\mathop{\arg\max}_{j}  (F_{j})
\end{equation}

\section{Result}
We test this method in public database DEAP\cite{RN300} and AMIGOS\cite{RN321}.

The proposed model is implemented by using the Tensorflow framework \cite{RN586} and deployed on NVIDIA Tesla K40c. The learning rate is set to 1E-3 with Adam Optimizer, and the keep probability of dropout operation is 0.5. The batch size for training and testing is set to 240. We use 10-fold cross-validation to evaluate the performance of our model. The average accuracy of the 10-fold validation processes is taken as the final result.

\subsection{Datasets and Pre-processing}
The DEAP \cite{RN300} is an open dataset for researchers to validate their model. This dataset contains 32 channels EEG signals and 8 channels peripheral physiological signals which are collected when 32 participants watched 40 videos each with one- minute duration. The EEG channels contain Fp1, AF3, F3, F7, FC5, FC1, C3, T7, CP5, CP1, P3, P7, PO3, O1, Oz, Pz, Fp2, AF4, Fz, F4, F8, FC6, FC2, Cz, C4, T8, CP6, CP2, P4, P8, PO4 and O2. The peripheral physiological channels contain hEOG, vEOG, zEMG, tEMG, GSR, Respiration belt (RB), Plethysmograph (Plethy), and skin temperature (Temp). Each trial contains 63s signals and the first 3s are the baseline signals. The baseline signals are recorded when the participants were under no stimulus. After watching a minute video, the participants rated a self-assessment of arousal, valence, liking, and dominance on a scale from 1 to 9. A preprocessed version had been provided: The data was down-sampled from 512Hz to 128Hz, and a bandpass frequency filter from 4.0-45.0Hz was applied.

\begin{figure}
\begin{minipage}[b]{.48\linewidth}
  \centering
  \centerline{\includegraphics[width=4.0cm]{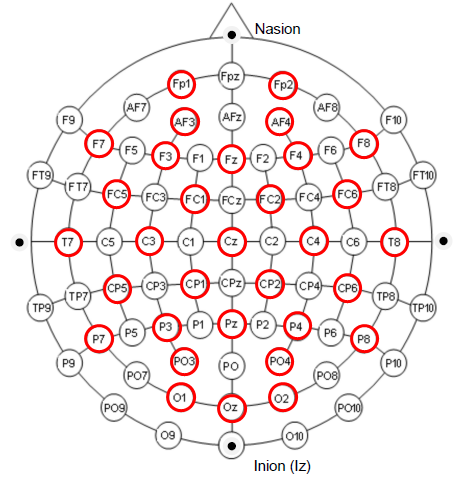}}
  \centerline{(a) 10-20 System}\medskip
\end{minipage}
\hfill
\begin{minipage}[b]{0.48\linewidth}
  \centering
  \centerline{\includegraphics[width=4.0cm]{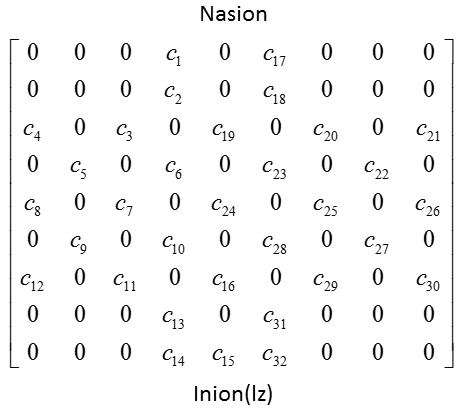}}
  \centerline{(b) 2D Electrode Topological Structure}\medskip
\end{minipage}
\caption{(a) The international 10-20 system describes the location of scalp electrodes, and the red nodes show the 32 electrodes used in DEAP dataset. (b) The 32 channels EEG signals are mapped into a 9x9 matrixes.}
\label{2D_DEAP}
\end{figure}

In the process of data pre-processing for one trial signals (a 40x8064 matrix, the row of the matrix contains 32 channels EEG signals and 8 channels peripheral physiological signals, and the column of the matrix contains 3s baseline and 60s stimulation signals with 128Hz) in DEAP dataset, the baseline signals (a 40x384 matrix) have been cut into 3 segments (each segment is a 40x128 matrix), and the mean value (a 40x128 matrix) of the baseline signals are calculated. And the data without baseline signals are cut into 60 segments (each segment is a 40x128 matrix) then minus the baseline signals mean value to get the preprocessed signals (a 40x7680 matrix, and the column of the matrix contains 60s signals with 128Hz). For each time sampling point, the 32 channels EEG signals are mapped into a 9x9 matrix (as shown in Fig.\ref{2D_DEAP}), getting the 2D electrode topological structure (a 9x9x7680 matrix) with Z-score normalization. Finally, the signals are cut into 60 segments with 1s length (each segment is a 9x9x128 matrix), and the 1s length was reported as the most suitable time window length in \cite{RN587}. In addition, we compare the High Scale CNN's performance with four different length time windows of 0.5s, 1s, 2s, 3s, and the result is shown in Table \ref{Time}. It is easy to see, the High Scale CNN model achieves the highest accuracy with the 1s length time windows. So we set the length of the segment to 1 second based on the model's performance. The final data size of EEG signals after processing is 76800 instances (the 76800 is calculated by 32 participants$\times$40 trails$\times$60s), and the dimension of each instance is 9x9x128. The 8 channels peripheral physiological signals are cut into 60 segments with 1s length in each modality, and the final data size of peripheral physiological signals in each modality after processing is 76800 instances, and the dimension of each instance is 128x1. In the DEAP dataset, the preprocessed data is randomly partitioned into 10 equal-size subsamples, and each subsample contains 7680 instances. So, there are 69120 instances in the training set, and 7680 instances in the test set in each fold of validation processes.

\begin{table}[]
\centering
\caption{Classification accuracies with different length time windows}
\label{Time}
\renewcommand{\arraystretch}{1.5} 
\begin{tabular}{|c|c|c|c|c|}
\hline
Length (s)    & 0.5   & 1     & 2     & 3     \\ \hline
Accuracy (\%) & 83.36 & 93.53 & 90.51 & 90.67 \\ \hline
\end{tabular}
\end{table}

The AMIGOS \cite{RN321} is a new open dataset. This dataset contains 14 channels EEG signals and 3 channels peripheral physiological signals which are collected when 40 participants watched 20 videos (16 short videos + 4 long videos). The EEG channels contain AF3, F7, F3, FC5, T7, P7, O1, O2, P8, T8, FC6, F4, F8, and AF4. The peripheral physiological channels contain ECG Right, ECG Left, and GSR. Each trial contains 5s baseline signals in first and the signals depend on the duration of the video. The baseline signals are recorded when the participants were under no stimulus. After watching the video, the participants rated a self-assessment of arousal, valence, liking, and dominance on a scale from 1 to 9. A preprocessed version had been provided: The data was down-sampled to 128Hz, and a bandpass frequency filter from 4.0-45.0Hz was applied.

\begin{figure}
\begin{minipage}[b]{.48\linewidth}
  \centering
  \centerline{\includegraphics[width=4.0cm]{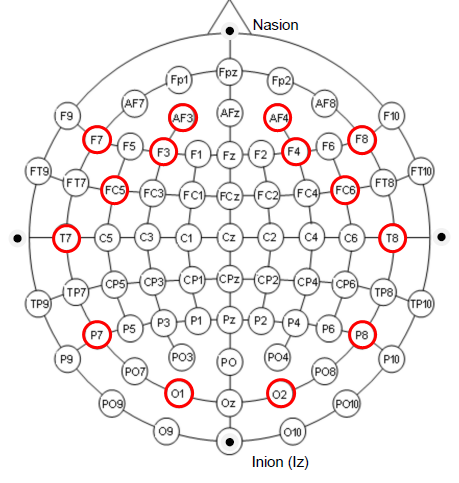}}
  \centerline{(a) 10-20 System}\medskip
\end{minipage}
\hfill
\begin{minipage}[b]{0.48\linewidth}
  \centering
  \centerline{\includegraphics[width=4.0cm]{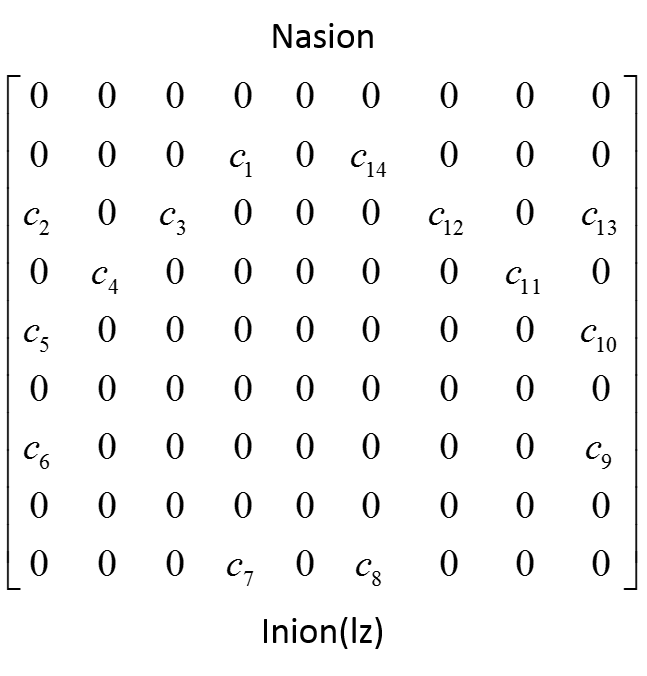}}
  \centerline{(b) 2D Electrode Topological Structure}\medskip
\end{minipage}
\caption{(a) The international 10-20 system describes the location of scalp electrodes, and the red nodes show the 14 electrodes used in AMIGOS dataset. (b) The 14 channels EEG signals are mapped into a 9x9 matrixes.}
\label{2D_AMIGOS}
\end{figure}

Here we use the signals which were recorded in short videos experiment. The participant ID of 9, 12, 21, 22, 23, 24 and 33 has been removed because there are some invalid data in the preprocessed version. The data pre-processing in the AMIGOS dataset is the same as in the DEAP dataset, and the signals also are segmented with 1s length. The 14 channels EEG signals are mapped into a 9x9 matrix (as shown in Fig.\ref{2D_AMIGOS}). The final data size of EEG signals after processing is 45474 9x9x128 matrices. The 3 channels peripheral physiological signals cut into 60 segments with 1s length in each modality, and the final data size of peripheral physiological signals in each modality after processing is 45474 128x1 matrices. In the AMIGOS dataset, the preprocessed data is randomly partitioned into 10 subsamples. One of the subsamples contains 4434 instances, and each of the remaining nine subsamples contains 4560 instances. So, there are 40914 instances in the training set and 4560 instances in the test set for the first nine rounds of validation processes, and there are 41040 instances in the training set and 4434 instances in the test set for the last round of validation process. 

\begin{figure}
\centering
\includegraphics[width=7cm]{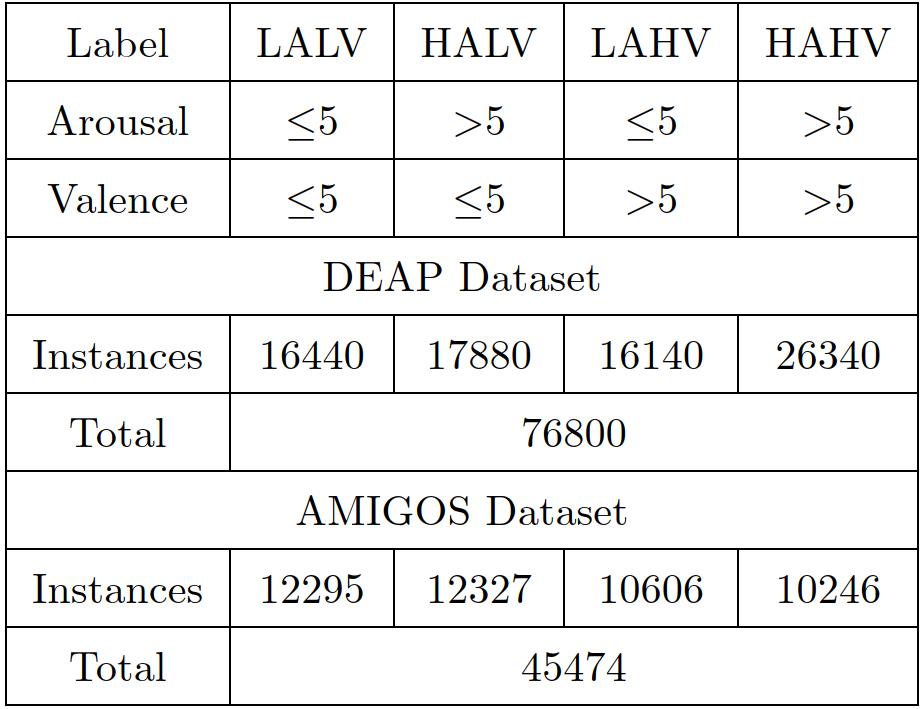}
\caption{Corresponding instance numbers in DEAP and AMIGOS Datasets.}
\label{Ins}
\end{figure}

Both datasets could be segmented in four classes: low arousal low valence (LALV), high arousal low valence (HALV), low arousal high valence (LAHV), high arousal high valence (HAHV) based on the arousal and valence value with the threshold of 5 respectively, and the corresponding instance numbers are shown in Figure \ref{Ins}.

%\begin{table}[]
%\centering
%\caption{Corresponding instance numbers in DEAP and AMIGOS Datasets}
%\label{Ins}
%\renewcommand{\arraystretch}{1.5} 
%\begin{tabular}{|c|c|c|c|c|}
%\hline
%Label    & LALV  & HALV            & LAHV            & HAHV            \\ \hline
%Arousal  & ≤$\le$5    & \textgreater{}5 & ≤$\le$5              & \textgreater{}5 \\ \hline
%Valence  & ≤$\le$5    & ≤$\le$5              & \textgreater{}5 & \textgreater{}5 \\ \hline
%\multicolumn{5}{|c|}{DEAP Dataset}        \\ \hline
%Instances & 16440 & 17880           & 16140           & 26340           \\ \hline
%Total    & \multicolumn{4}{c|}{76800}                                  \\ \hline
%\multicolumn{5}{|c|}{AMIGOS Dataset}        \\ \hline
%Instances & 12295 & 12327           & 10606           & 10246           \\ \hline
%Total    & \multicolumn{4}{c|}{45474}                               \\ \hline
%\end{tabular}
%\end{table}

\begin{figure}
\centering
\includegraphics[width=7cm]{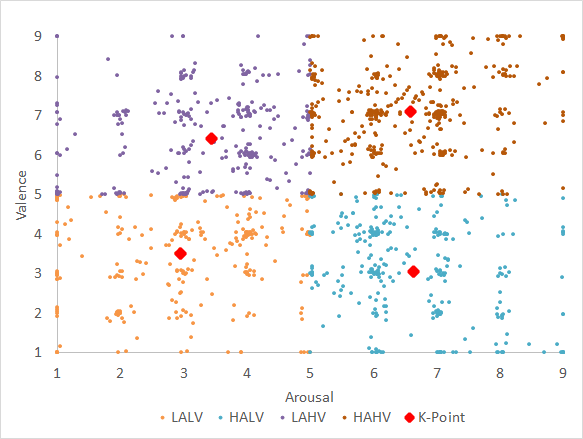}
\caption{Distribution of affective states categories in DEAP, and the red points are the mean value of the corresponding labels on Arousal-Valence data, such as LALV (2.95, 3.51), HALV (6.64, 3.07), LAHV (3.44, 6.42), HAHV (6.58, 7.11).}
\label{PDEAP}
\end{figure}

\begin{figure}
\centering
\includegraphics[width=7cm]{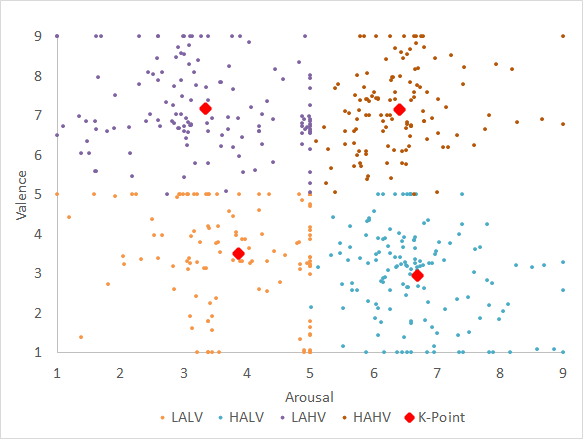}
\caption{Distribution of affective states categories in AMIGOS, and the red points are the mean value of the corresponding labels on Arousal-Valence data, such as as LALV (3.86, 3.52), HALV (6.69, 2.95), LAHV (3.35, 7.18), HAHV (6.42, 7.17).}
\label{PAMIGOS}
\end{figure}

Fig.\ref{PDEAP} and Fig.\ref{PAMIGOS} show the distribution of affective states categories in DEAP and AMIGOS datasets respectively. The red points are calculated by the mean value of the corresponding category labels on Arousal-Valence data. It could represent the categories in two-dimensional space such as the LALV (2.95, 3.51), HALV (6.64, 3.07), LAHV (3.44, 6.42), HAHV (6.58, 7.11) in DEAP dataset, and LALV (3.86, 3.52), HALV (6.69, 2.95), LAHV (3.35, 7.18), HAHV (6.42, 7.17) in AMIGOS dataset. Then we calculate the Euclidean distance between various category labels and using the standard normal distribution to calculate the classification score reliability. In each modality, we calculate the final classification score through classification score from CNN model and classification score reliability, and use the standard deviation of final classification score to represent the corresponding modality reliability, then calculate the result of multimodal information integration.

\subsection{Fusion Result}
\begin{figure}
\centering
\includegraphics[width=8cm]{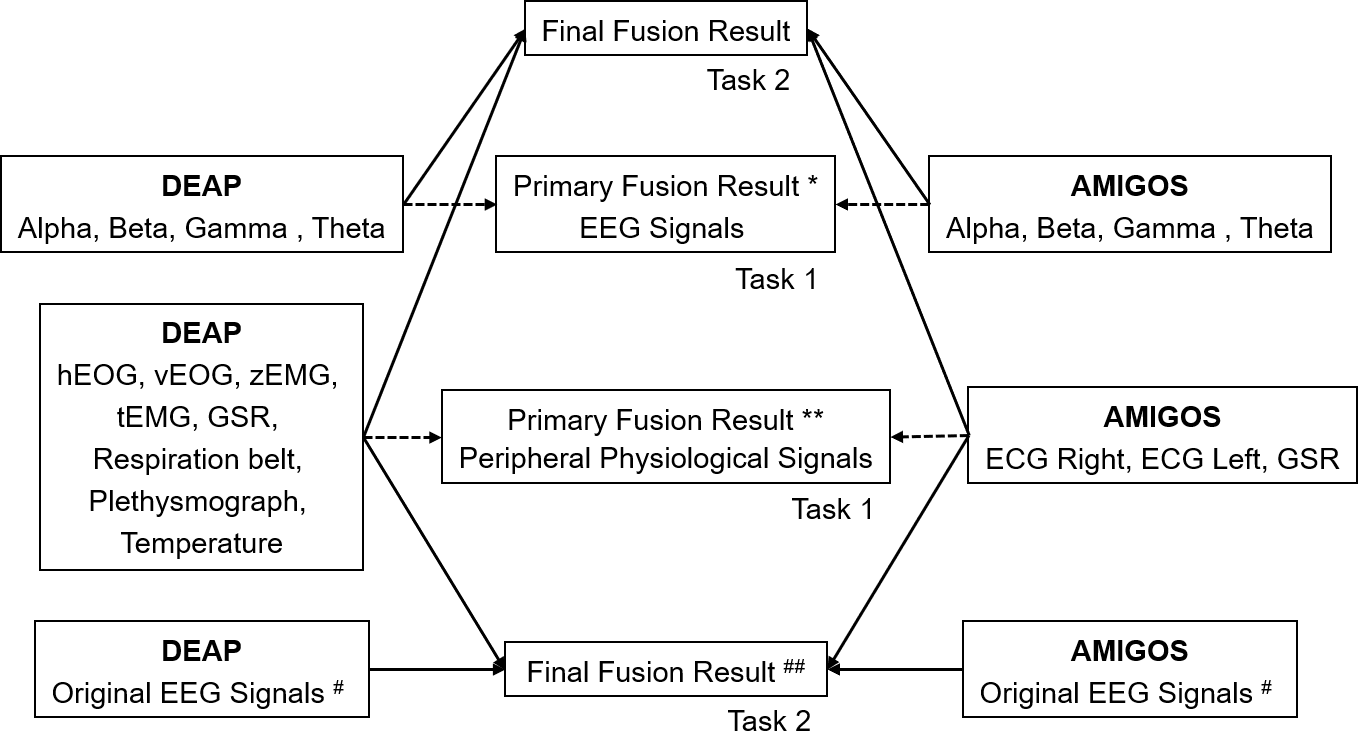}
\caption{Two tasks in the fusion performance analysis.}
\label{Task}
\end{figure}

We use 2 tasks to verify the fusion performance of the biologically inspired decision fusion model: the comparison of single-modality results with the primary fusion results (Task 1) and the comparison of primary fusion results with the final fusion result (Task 2) as shown in Fig.\ref{Task}. The primary fusion results are the fusion results of EEG signals and peripheral physiological signals respectively, and the final fusion result is the result that combines all the EEG signals and peripheral physiological signals. In addition, to verify the multimodal fusion model of EEG signals, the signals are decomposed into four parts according to the four frequency bands of theta (4-7 Hz), alpha (8-13 Hz), beta (14-30 Hz) and gamma (31-45 Hz). Besides, considering that voting methods are widely used in other decision fusion studies\cite{RN629,RN679}, here we compare the performance of the plurality voting method and the biologically inspired decision fusion model on the peripheral physiological signals in DEAP dataset (Task 3).

Table \ref{Task1T} shows the mean accuracy rate and the standard deviations of 10 folds cross-validation of each single-modality and the increment in primary fusion result (Task 1). It is easy to see, there is a significant increase in the accuracy rate in the primary fusion result compared with the single-modality results. In the DEAP dataset, compared with the result in single-modality, the fusion model improves the accuracy by 6-22\% in EEG signals, and 10-45\% in peripheral physiological signals. In the AMIGOS dataset, the fusion model improves the accuracy by 4-15\% in EEG signals, and 0.03-15\% in peripheral physiological signals. And in ECG signals, the improvement of accuracy is not significant due to the high recognition accuracy in single-modality.

\begin{table}[]
\centering
\caption{Accuracy improvement in primary fusion result of EEG signals and peripheral physiological signals compared with the single-modality result (Task 1)}
\label{Task1T}
\renewcommand{\arraystretch}{1.5} 
\begin{threeparttable}
\begin{tabular}{|c|c|c|}
\hline
Modality         & Single-modality Result (\%) & \begin{tabular}[c]{@{}c@{}}Improvement in\\ Fusion Result (\%)\end{tabular} \\ \hline
\multicolumn{3}{|c|}{\textbf{EEG Signals (DEAP)}}                                                                              \\ \hline
Alpha            & 89.01 $\pm$ 0.22           & 6.76                                                                 \\ \hline
Beta             & 89.69 $\pm$ 1.01         & 6.08                                                                 \\ \hline
Gamma            & 73.09 $\pm$ 1.12            & 22.68                                                               \\ \hline
Theta            & 81.97 $\pm$ 0.27         & 13.80                                                              \\ \hline
Primary Fusion\tnote{*}            & \multicolumn{2}{c|}{95.77 $\pm$ 0.4}                                                               \\ \hline
\multicolumn{3}{|c|}{\textbf{Peripheral Physiological Signals (DEAP)}}                                                         \\ \hline
hEOG             & 72.57 $\pm$ 1.28        & 24.70                                                               \\ \hline
vEOG             & 86.45 $\pm$ 0.81           & 10.82                                                               \\ \hline
zEMG             & 73.66 $\pm$ 0.61           & 23.61                                                               \\ \hline
tEMG             & 83.74 $\pm$ 1.53           & 13.53                                                                \\ \hline
GSR              & 74.60 $\pm$ 0.35           & 22.67                                                                \\ \hline
Respiration belt & 66.33 $\pm$ 2.48          & 30.94                                                                \\ \hline
Plethysmograph   & 72.94 $\pm$ 2.23          & 24.33                                                               \\ \hline
Temperature      & 51.43 $\pm$ 0.46         & 45.84                                                                \\ \hline
Primary Fusion\tnote{**}             & \multicolumn{2}{c|}{97.27 $\pm$ 0.27}                                                             \\ \hline
\multicolumn{3}{|c|}{\textbf{EEG Signals (AMIGOS)}}                                                                                                                    \\ \hline
Alpha     & 79.54 $\pm$ 0.91                                                       & 11.53                                                              \\ \hline
Beta      & 86.80 $\pm$ 0.16                                                       & 4.27                                                                \\ \hline
Gamma     & 85.39 $\pm$ 0.21                                                       & 5.68                                                               \\ \hline
Theta     & 75.88 $\pm$ 0.24                                                        & 15.19                                                               \\ \hline
Primary Fusion\tnote{*}            & \multicolumn{2}{c|}{91.07 $\pm$ 0.66}                                                               \\ \hline
\multicolumn{3}{|c|}{\textbf{Peripheral Physiological Signals (AMIGOS)}}                                                                                               \\ \hline
ECG Right & 99.71 $\pm$ 0.08                                                      & 0.03                                                                \\ \hline
ECG Left  & 99.30 $\pm$ 0.17                                                      & 0.44                                                                \\ \hline
GSR       & 84.69 $\pm$ 1.18                                                     & 15.05                                                                \\ \hline
Primary Fusion\tnote{**}             & \multicolumn{2}{c|}{99.74 $\pm$ 0.12}                                                             \\ \hline
\end{tabular}
\begin{tablenotes}
        \footnotesize
        \item[*] The fusion result of four frequency bands EEG signals
        \item[**] The fusion result of 8 multimodal peripheral physiological signals
      \end{tablenotes}
\end{threeparttable}
\end{table}

Table \ref{Task2T} shows the comparison of primary fusion results with the final fusion result (Task 2). It is easy to see, the accuracy rate will rise further with the fusion between EEG signals and peripheral physiological signals. We also use the original EEG signals to verify the method's performance in the final fusion. Compared with the primary fusion results of EEG signals and the original EEG signals, the final fusion result shows a 3-8\% improvement in  DEAP and AMIGOS datasets. Compared with the primary fusion results of peripheral physiological signals, the improvement of accuracy in final fusion result is not significant. The comparison of single-modality with primary fusion results and the final fusion result could be found in supplemental material Table S1 and Table S2.

\begin{table}[]
\centering
\caption{Accuracy improvement in final fusion result compared with the primary fusion result (Task 2)}
\label{Task2T}
\renewcommand{\arraystretch}{1.5} 
\begin{threeparttable}
\begin{tabular}{|c|c|c|c|}
\hline
\multirow{2}{*}{Modality} & \multirow{2}{*}{Result} & \multicolumn{2}{c|}{Improvement in Final Fusion} \\ \cline{3-4} 
                          &                         & Final Fusion           & Final Fusion\tnote{\#\#}           \\ \hline
\multicolumn{4}{|c|}{\textbf{DEAP}}                                                                    \\ \hline
EEG\tnote{\#}                      & 93.53 $\pm$ 0.36                 & 5.64                   & 4.99                    \\ \hline
Primary Fusion EEG        & 95.77 $\pm$ 0.4                 & 3.4                    & 2.75                    \\ \hline
Primary Fusion Peripheral & 97.27 $\pm$ 0.27                 & 1.9                    & 1.25                    \\ \hline
Final Fusion              & \multicolumn{3}{c|}{99.17 $\pm$ 0.03}                                                 \\ \hline
Final Fusion\tnote{\#\#}             & \multicolumn{3}{c|}{98.52 $\pm$ 0.09}                                                 \\ \hline
\multicolumn{4}{|c|}{\textbf{AMIGOS}}                                                                  \\ \hline
EEG\tnote{\#}                      & 95.86 $\pm$ 0.34                 & 3.18                   & 4.03                    \\ \hline
Primary Fusion EEG        & 91.07 $\pm$ 0.66                 & 7.97                   & 8.82                    \\ \hline
Primary Fusion Peripheral & 99.74 $\pm$ 0.12                  & -0.7                   & 0.15                    \\ \hline
Final Fusion              & \multicolumn{3}{c|}{99.04 $\pm$ 0.09}                                                 \\ \hline
Final Fusion\tnote{\#\#}             & \multicolumn{3}{c|}{99.89 $\pm$ 0.05}                                                 \\ \hline
\end{tabular}
\begin{tablenotes}
        \footnotesize
        \item[\#] The result in original EEG signals
        \item[\#\#] The final fusion result of peripheral physiological signals and original EEG signals
      \end{tablenotes}
\end{threeparttable}
\end{table}

We calculate the precision, sensitivity, specificity and f-measure by the Equation~\ref{p-p} -~\ref{p-f}, where TP, TN, FP, and FN refer respectively to ``True Positives", ``True Negatives", ``False Positives" and ``False Negatives" respectively. And the result of Final Fusion$^{\#\#}$ which represents the final fusion result of peripheral physiological signals and original EEG signal in DEAP and AMIGOS dataset is shown in Table \ref{Performance}. It can be seen that the proposed model makes a good performance in per valence-arousal quadrant. The performance metrics for single-modality could be found in supplemental material Table S3 - S6.

\begin{equation}
\label{p-p}
Precision = \frac{TP}{TP+FP}
\end{equation}

\begin{equation}
\label{p-se}
Sensitivity = \frac{TP}{TP+FN}
\end{equation}

\begin{equation}
\label{p-sp}
Specificity = \frac{TN}{TN+FP}
\end{equation}

\begin{equation}
\label{p-f}
F1 = \frac{2*Precision*Sensitivity}{Precision+Sensitivity}
\end{equation}

\begin{table}[]
\centering
\caption{Performance metrics for Final Fusion$^{\#\#}$ in DEAP and AMIGOS dataset}
\label{Performance}
\renewcommand{\arraystretch}{1.5} 
\begin{tabular}{|c|c|c|c|c|}
\hline
Label         & Precision      & Sensitivity    & Specificity    & F1             \\ \hline
\multicolumn{5}{|c|}{DEAP Dataset}                                                \\ \hline
LALV          & 99.44          & 97.93          & 99.85          & 98.68          \\ \hline
HALV          & 99.10          & 98.23          & 99.73          & 98.66          \\ \hline
LAHV          & 99.49          & 97.36          & 99.87          & 98.41          \\ \hline
HAHV          & 97.02          & 99.77          & 98.39          & 98.38          \\ \hline
\textbf{Mean} & \textbf{98.76} & \textbf{98.32} & \textbf{99.46} & \textbf{98.53} \\ \hline
\multicolumn{5}{|c|}{AMIGOS Dataset}                                              \\ \hline
LALV          & 99.91          & 99.83          & 99.97          & 99.87          \\ \hline
HALV          & 99.68          & 99.84          & 99.88          & 99.76          \\ \hline
LAHV          & 99.91          & 99.78          & 99.97          & 99.84          \\ \hline
HAHV          & 99.81          & 99.86          & 99.94          & 99.84          \\ \hline
\textbf{Mean} & \textbf{99.83} & \textbf{99.83} & \textbf{99.94} & \textbf{99.83} \\ \hline
\end{tabular}
\end{table}

%%%%%%% 
We use lock box approach \cite{HOSSEINI2020456} to determine whether overhyping has occurred in the Multiscale CNNs models. The lock box approach is a new technique that could be used to determine whether overhyping has occurred. The data is first divided into a hyperparameter optimization set and a lock box. And the lock box only be accessed just one time to generate an unbiased estimate of the model's performance. In the DEAP and AMIGOS datasets, 90\% of the data are set aside in hyperparameter optimization set and the remaining 10\% of the data set aside in a lock box. With the 10-fold cross-validation approach, the hyperparameters in Multiscale CNNs models can be iteratively modified on the hyperparameter optimization set. When the average accuracy in the model is good enough, the model be tested on the lock box data. Compared with the training result on the hyperparameter optimization set, the testing result on the lock box data did not decrease significantly. It can prove that the overhyping has not occurred in the Multiscale CNNs. The result is shown in Table \ref{Lockbox}.

\begin{table}[]
\centering
\caption{The result of lock box approach. The training result on the hyperparameter optimization set and the testing result on the lock box set.}
\label{Lockbox}
\renewcommand{\arraystretch}{1.5} 
\begin{tabular}{|c|c|c|}
\hline
%Modality            & Training Result (\%)    & Testing Result (\%)   \\ \hline
Modality            & Train (\%)    & Test (\%)   \\ \hline
\multicolumn{3}{|c|}{EEG Signals (DEAP)}                        \\ \hline
Alpha               & 89.70                & 89.18              \\ \hline
Beta                & 89.92                & 90.41              \\ \hline
Gamma               & 73.94                & 73.83              \\ \hline
Theta               & 81.15                & 81.03              \\ \hline
EEG\#               & 93.82                & 93.96              \\ \hline
\multicolumn{3}{|c|}{Peripheral Physiological Signals (DEAP)}   \\ \hline
hEOG                & 73.26                & 73.03              \\ \hline
vEOG                & 85.92                & 85.73              \\ \hline
zEMG                & 73.50                & 72.54              \\ \hline
tEMG                & 84.87                & 84.27              \\ \hline
GSR                 & 73.15                & 71.39              \\ \hline
Respiration belt    & 69.23                & 68.45              \\ \hline
Plethysmograph      & 74.03                & 74.08              \\ \hline
Temperature         & 51.98                & 52.04              \\ \hline
\multicolumn{3}{|c|}{EEG Signals (AMIGOS)}                      \\ \hline
Alpha               & 81.21                & 81.46              \\ \hline
Beta                & 87.32                & 87.62              \\ \hline
Gamma               & 85.64                & 84.79              \\ \hline
Theta               & 77.35                & 76.87              \\ \hline
EEG\#               & 96.50                & 96.14              \\ \hline
\multicolumn{3}{|c|}{Peripheral Physiological Signals (AMIGOS)} \\ \hline
ECG Right           & 99.67                & 99.75              \\ \hline
ECG Left            & 99.48                & 99.31              \\ \hline
GSR                 & 85.64                & 86.82              \\ \hline
\end{tabular}
\end{table}

\subsection{Fusion Performance Analysis}

\begin{figure}
\centering
\includegraphics[width=7cm]{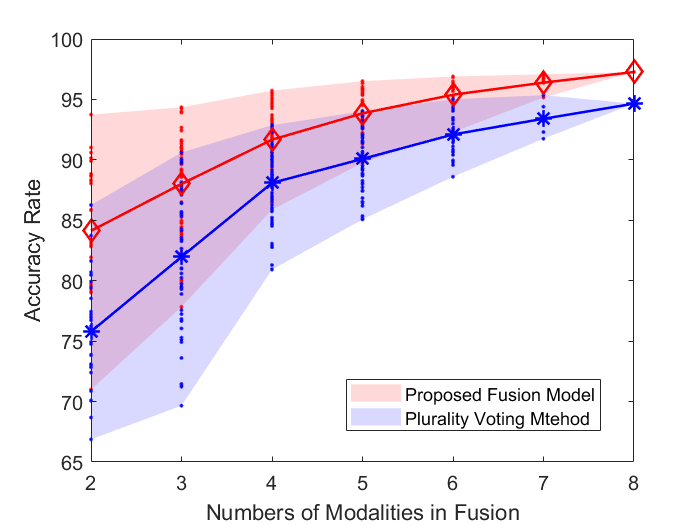}
\caption{The comparison of the plurality voting method with the biologically inspired decision fusion model (Task 3). The red $\diamondsuit $ and the blue $*$ represent the average accuracy for different numbers of  modalities in the biologically inspired decision fusion model and plurality voting method respectively.}
\label{Multi}
\end{figure}

There are 8 modalities peripheral physiological signals in the DEAP dataset which could be used to verify the performance of the fusion model. The comparison of the plurality voting method with the biologically inspired decision fusion model (Task 3) is shown in Fig.\ref{Multi} and Table \ref{MultiData}. The plurality voting method takes the class label which receives the largest number of votes as the final result. In Fig.\ref{Multi}, the red and blue dots respent the results based on biologically inspired decision fusion model and plurality voting method respectively. There are 247 results in different combinations of various modalities in each method. The red $\diamondsuit $ and the blue $*$ represent the average accuracy for different numbers of  modalities in each method. It is easy to see, the accuracy rate can be improved more effectively by the biologically inspired decision fusion model than by using plurality voting, especially when the number of modalities is small. Compared with the nondiscriminatory plurality voting method, the biologically inspired decision fusion model considers the correlation between various class labels. Then the reliability of each single-modality signal is calculated by this correlation and the classification probability from physiological signals. Finally, the decision is made by the more reliable modality information while it retains other modalities information. The best modality combination in different modality numbers could be found in supplemental material Table S7. And we use a sample of peripheral physiological signals fusion in the DEAP dataset to show the details of how to calculate the $d_{ij}$, $f(d_{ij})$, $GauPR_{m,j}$, $S_{m}$, and $F_{j}$ in the supplemental material.

Table \ref{MultiData} shows the maximum, minimum, mean values and the standard deviation of fusion results with different modality numbers in the biologically inspired decision fusion model and plurality voting method respectively. With the increase of the modality numbers, the accuracy of the proposed fusion model increases and achieves the highest accuracy when the modality number is 8, and the accuracy of the plurality voting method achieves the highest accuracy when the modality number is 7. Compared with the plurality voting method, the proposed fusion model can improve the minimum accuracy more effectively - about 4\% increment in different modality numbers. In other words, the proposed fusion model can effectively utilize modalities with poor classification performance. For example, without the vEOG and tEMG which achieve the high accuracies in single-modality result, the proposed fusion model achieves the accuracy of 92.29\%. Without the vEOG which achieves the highest accuracies in single-modality result, the proposed fusion model achieves the accuracy of 95.16\% - about 2\% below the maximum value of 97.08\%. The mean values and the standard deviation of both methods indicate that the increase of the modality numbers help to improve the recognition accuracy and the robustness.

\begin{table}[]
\centering
\caption{The fusion result of different modality numbers in peripheral physiological signals (\%)}
\label{MultiData}
\renewcommand{\arraystretch}{1.5}
\begin{threeparttable} 
\begin{tabular}{|c|c|c|c|c|c|c|c|c|}
\hline
\multirow{2}{*}{\begin{tabular}[c]{@{}c@{}}M\end{tabular}} & \multicolumn{4}{c|}{Proposed Fusion Model} & \multicolumn{4}{c|}{Plurality Voting Method} \\ \cline{2-9} 
                                                                              & Max             & Min            & Mean           & SD            & Max       & Min       & Mean      & SD       \\ \hline
2                                                                             & 93.74           & 70.99          & 84.13          & 5.34          & 86.26     & 66.87     & 75.79     & 4.48     \\ \hline
3                                                                             & 94.35           & 77.84          & 88.01          & 3.77          & 90.61     & 69.66     & 81.98     & 5.20     \\ \hline
4                                                                             & 95.73           & 85.87          & 91.67          & 2.49          & 92.87     & 80.92     & 88.09     & 2.97     \\ \hline
5                                                                             & 96.51           & 89.75          & 93.85          & 1.73          & 94.02     & 85.07     & 90.08     & 2.35     \\ \hline
6                                                                             & 96.90           & 92.29          & 95.39          & 1.10          & 95.02     & 88.60     & 92.10     & 1.62     \\ \hline
7                                                                             & 97.08           & 95.16          & 96.39          & 0.66          & 95.35     & 91.75     & 93.40     & 1.05     \\ \hline
8                                                                             & 97.27           & $\backslash$         & $\backslash$          & $\backslash$          & 94.65     & $\backslash$     & $\backslash$     & $\backslash$     \\ \hline
\end{tabular}
\begin{tablenotes}
        \footnotesize
        \item M is the numbers of modalities used in the biologically inspired decision fusion model
	\end{tablenotes}
\end{threeparttable}
\end{table}

%We use the primary fusion result which combined 8 multimodal peripheral physiological signals in the DEAP dataset to analyze the proposed fusion model. 
In the proposed fusion model, we assume that: (1) for the samples which predicted fusion results are true, there is a strong correlation between final classification score of the true label and the modality reliability; (2) for the samples which predicted fusion results are false, there is a weak correlation between final classification score of the true label and the modality reliability. So We use the cosine similarity to calculate the correlation between the final classification score and the modality reliability in one sample as shown in Equation~\ref{cos}. In Equation~\ref{cos}, the $GauPR_{m,true}$ is the final classification score of the true label in modality $m$, and the $S_{m}$ represents the modality reliability of modality $m$.

\begin{equation}
\label{cos}
\cos (\theta _{GauPR,S})=\frac{\sum_{m=1}^{8}GauPR_{m,true}\times S_{m}}{\sqrt{\sum_{m=1}^{8}GauPR_{m,true}^{2}}\times {\sqrt{\sum_{m=1}^{8}S_{m}^{2}}}}
\end{equation}

\begin{figure}
\centering
\includegraphics[width=7cm]{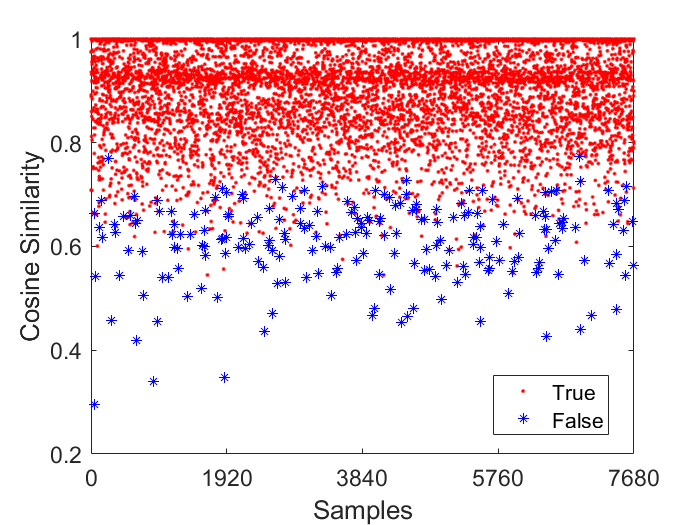}
\caption{The result of the test samples' cosine similarity in primary fusion. The red dots represent the predicted results which are true, and the blue $*$ represent the predicted results which are false.}
\label{CS}
\end{figure}

The result of the test samples' cosine similarity is shown in Fig.\ref{CS}. There are 7680 test samples in each round of 10-fold validation. The cosine similarity of more than 84\% samples are greater than 0.8, and more than 91\% samples are greater than 0.75. There are three cases when calculating the cosine similarity: (1) if the final classification score of the true label in one modality is high, the modality reliability of the correlation modality is high, so the cosine similarity is high which means the correlation is strong. (2) if the final classification scores of all labels in one modality are low, the modality reliability of the correlation modality is low, so the cosine similarity is high which means the correlation is strong. (3) if the final classification score of the true label in one modality is low while the score of one false label is high, the modality reliability of the correlation modality is high, so the cosine similarity is low which means the correlation is weak.

\begin{figure*}
\begin{minipage}[b]{.48\linewidth}
  \centering
  \centerline{\includegraphics[width=9.0cm]{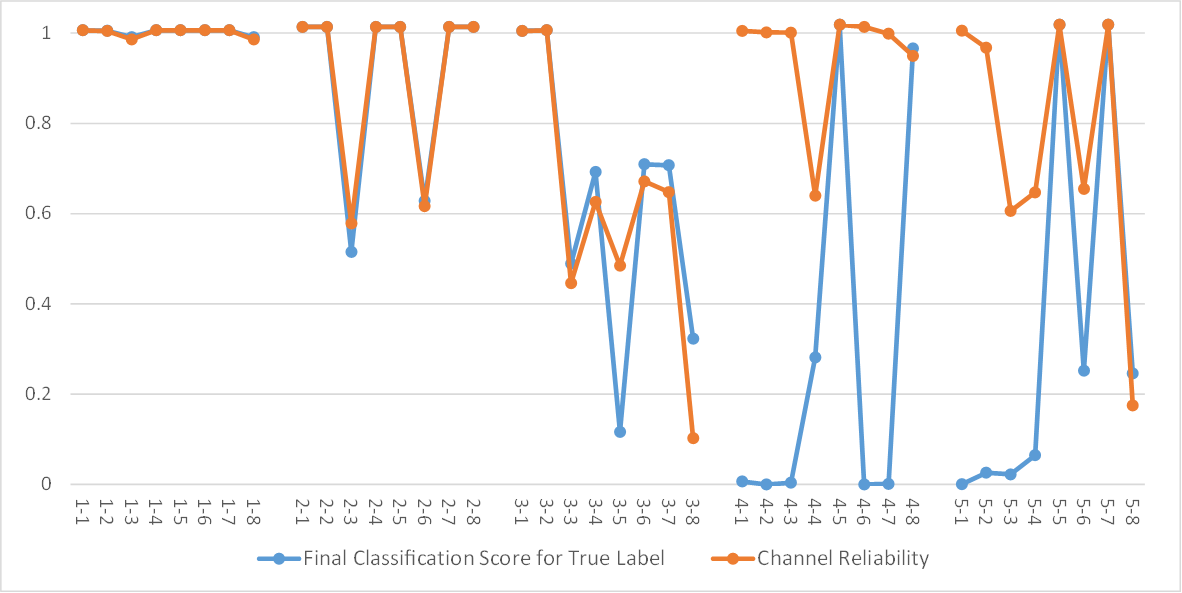}}
  \centerline{(a) Five samples}\medskip
\end{minipage}
\hfill
\begin{minipage}[b]{0.48\linewidth}
  \centering
  \centerline{\includegraphics[width=9.0cm]{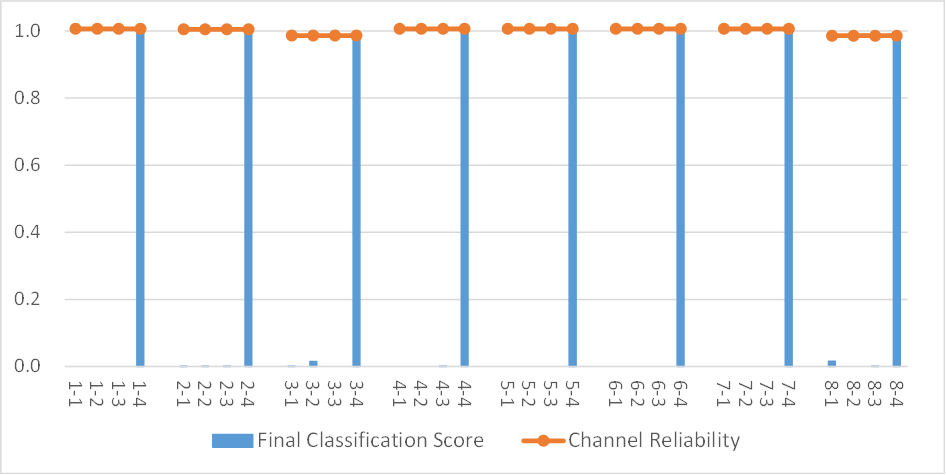}}
  \centerline{(b) Sample 1}\medskip
\end{minipage}
\hfill
\begin{minipage}[b]{0.48\linewidth}
  \centering
  \centerline{\includegraphics[width=9.0cm]{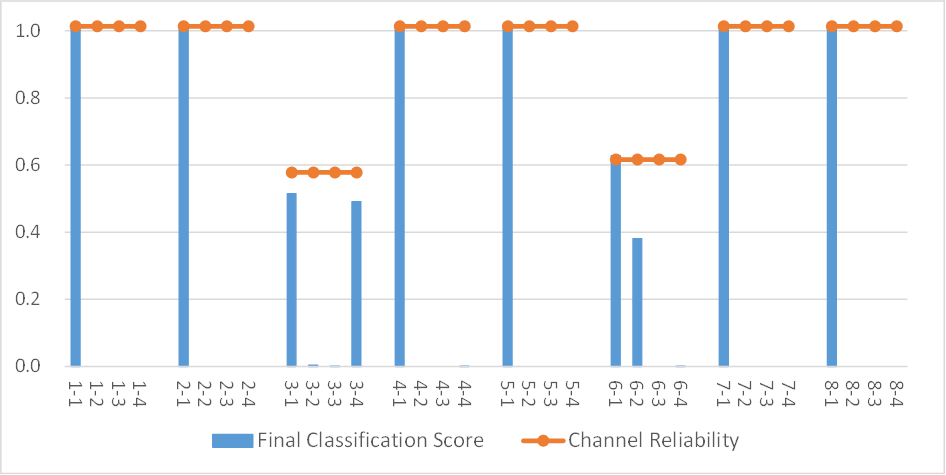}}
  \centerline{(c) Sample 2}\medskip
\end{minipage}
\hfill
\begin{minipage}[b]{0.48\linewidth}
  \centering
  \centerline{\includegraphics[width=9.0cm]{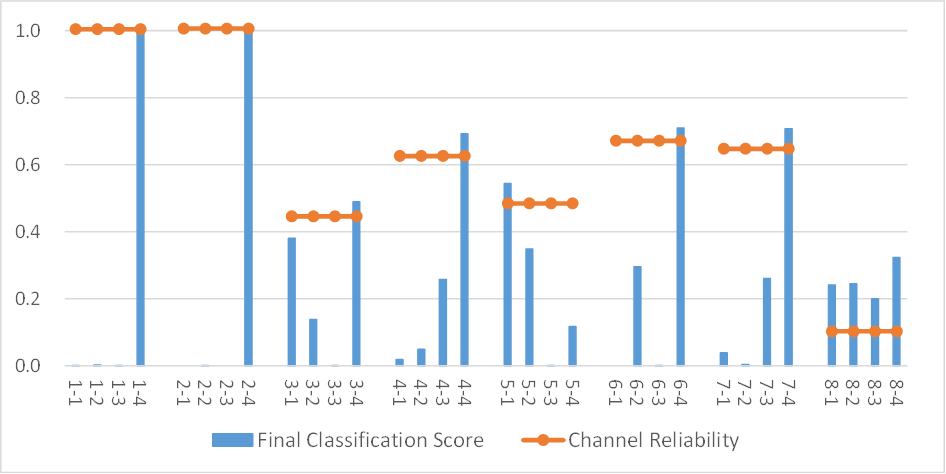}}
  \centerline{(d) Sample 3}\medskip
\end{minipage}
\hfill
\begin{minipage}[b]{0.48\linewidth}
  \centering
  \centerline{\includegraphics[width=9.0cm]{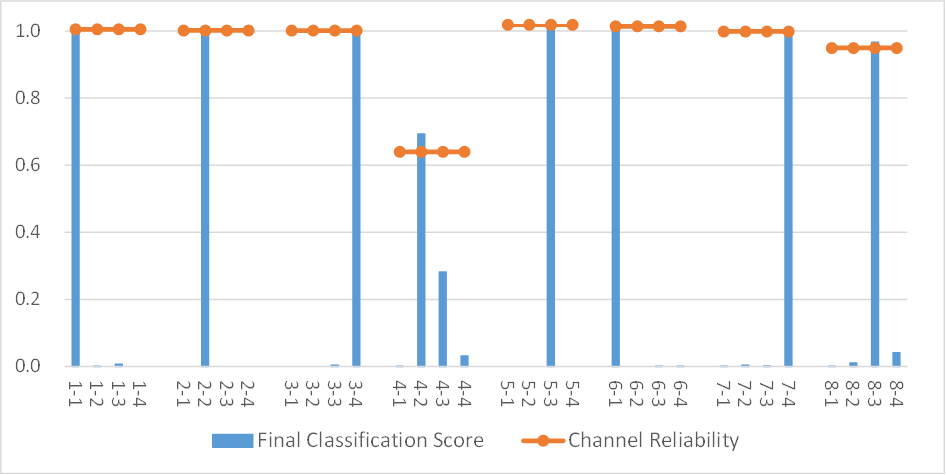}}
  \centerline{(e) Sample 4}\medskip
\end{minipage}
\hfill
\begin{minipage}[b]{0.48\linewidth}
  \centering
  \centerline{\includegraphics[width=9.0cm]{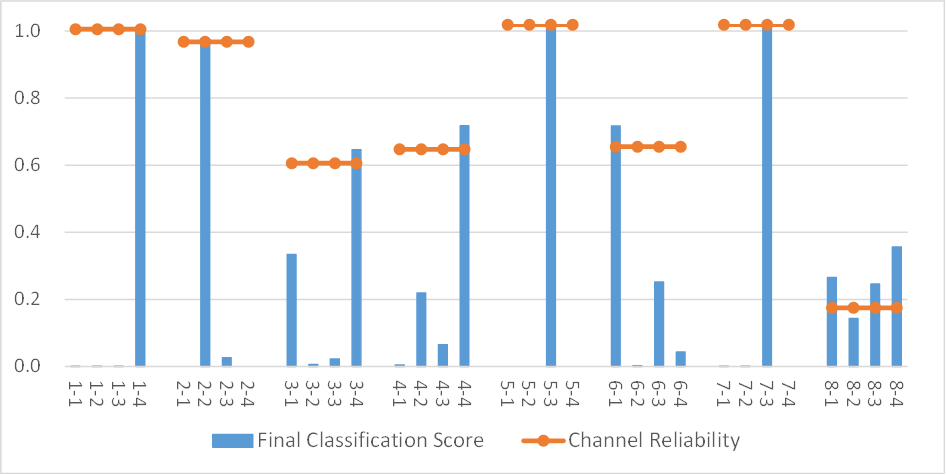}}
  \centerline{(f) Sample 5}\medskip
\end{minipage}
\caption{\textbf{(a)} Five samples. The blue line represents the result of the final classification score of the true label, and the orange line represents the correlation modality reliability. The horizontal axis is the sample's ID and it's modality, such as `1-8' means that the sample's ID is 1 and the modality is Temperature. There is a strong correlation between the final classification score of the true label and the modality reliability when the sample's ID is 1, 2, 3, and the cosine similarity is 1, 1, 0.97 respectively. The correlation is weak when the sample's ID is 4, 5, and the cosine similarity is 0.56, 0.69 respectively. In the rest of the figures, the horizontal axis is the modality ID and the label, such as `2-4' means that the modality is vEOG and the label is HAHV. 
\textbf{(b)} Sample 1. Sample 1 represents the case that all modalities predicted the true results and the final classification scores are high. For sample 1, the true label is 4 (HAHV). All modalities contribute strongly to the decision making. 
\textbf{(c)} Sample 2. Sample 2 represents the case that all modalities predicted the true results but the final classification scores the partial modalities are low. For sample 2, the true label is 1 (LALV). All modalities predicted the true results, but the final classification scores are low in modality 3 (zEMG) and 6 (Respiration belt), so the modality reliability of these modalities is low. Compared with others modality, these two modalities contribute weakly to decision making.
\textbf{(d)} Sample 3. Sample 3 represents the case that most of the modalities predicted the true results, and few modalities predicted the false results. For sample 3, the true label is 4 (HAHV). Only modality 5 (GSR) predicted the false result.
\textbf{(e)} Sample 4. Sample 4 represents the case that the numbers of modalities which predicted the true results are equal to the numbers of modalities which predicted the false results. For sample 4, the true label is 3 (LAHV). The number of modalities predicted as 1 (LALV), 2 (HALV), 3 (LAHV), 4 (HAHV) is equal. With the contribution of modality 4 (tEMG) which predicted LAHV with a low final classification score and medium modality reliability, the proposed fusion model could predict the true results.
\textbf{(f)} Sample 5. Sample 5 represents the case that few modalities predicted the true results, and most of the modalities predicted the false results. For sample 5, the true label is 3 (LAHV). Only two modalities 5 (GSR) and 7 (Plethysmograph) predicted as 3 (LAHV), and the four modalities 1 (hEOG), 3 (zEMG), 4 (tEMG), 8 (Temperature) predicted as 4 (HAHV). With the contribution of the modalities which predicted LAHV with a low or medium final classification score and the modality reliability, the proposed fusion model could predict the true results.}
\label{Samples}
\end{figure*}

In addition, we select 5 samples to show the details in the process of the proposed fusion model, as shown in Fig.\ref{Samples}. To make the final classification score and modality reliability on the same scale, we double the value of modality reliability. Fig. \ref{Samples} could prove that the proposed fusion model makes the decision by the more reliable modality information while other modalities information is retained.

\subsection{Comparison with Existing Models}

The compared methods used a variety of approaches. The affective states recognition methods could be simply divided into traditional machine learning approaches and deep learning approaches, and the fusion technologies mainly contain feature level fusion and decision level fusion.

The traditional machine learning approaches \cite{RN630,RN312,RN629} use well-designed classifiers with hand-crafted features which may be limited to domain knowledge. Motivated by the outstanding performance of deep learning approaches in pattern recognition tasks, more and more studies use deep learning approaches for deep feature extraction \cite{RN520,RN425,RN627} or affective states recognition directly \cite{RN678,RN679,RN486,RN677}. Compared with these affective states recognition methods, the Multiscale CNNs can better handle data from different dimensions: the High Scale CNN is used to process high-dimensional EEG signals, which retains the spatial and temporal information of the signals; the Low Scale CNN is used to process low-dimensional peripheral physiological signals, which retains more useful temporal information in the signals.

Most of the fusion technologies in these methods are feature level fusion. The disadvantage of feature level fusion is that it can not directly distinguish which type of modalities classification is the best, and can not directly obtain the fusion results of different modalities combinations. The fusion technologies in \cite{RN629,RN679,RN677} are decision level fusion, but most of them are simple voting methods. The disadvantage of the voting method is that only when the number of modalities is large, the fusion result is good. When the number of modalities is low, the fusion result is not improved significantly.

The Biologically Inspired Decision Fusion Model makes decisions by the more reliable modality information while other modalities information is retained. Compared with the feature level fusion technologies, the proposed model is more explanatory and can directly obtain the fusion results of different modalities combinations. Compared with the voting methods, the proposed model has a better improvement effect, and has achieved better fusion results when the number of modalities is low.

The existing multimodal affective states recognition studies used various EEG channels, peripheral physiological signals and labels. In the DEAP dataset, the signals data used in Shu et al.\cite{RN630}, Yin et al.\cite{RN312}, Ma et al.\cite{RN678} and Bagherzadeh et al.\cite{RN629} are consistent with us. In the AMIGOS dataset, the signals data used in Yang et al.\cite{RN683} and Li et al.\cite{RN677} are consistent with us. As a new dataset AMIGOS, the model which proposed in Shukla et al.\cite{RN682} and Harper et al.\cite{RN681} used single-mode signals, and other models used multimodal signals. It easy to see, the proposed model outperformed these other methods in terms of accuracy in the DEAP and AMIGOS datasets - the accuracy rate increase 4.92\% in the DEAP dataset and 9.89\% in the AMIGOS dataset. 

\begin{table*}[]
\scriptsize
\centering
\caption{The comparison of our model with previous multimodal affective states recognition studies}
\label{Compared}
\renewcommand{\arraystretch}{3.5} 
\begin{threeparttable}
\begin{tabular}{|c|c|c|c|c|}
\hline
Research & Year & Methods and Signals & Labels & Accuracy(\%) \\ \hline
\multicolumn{5}{|c|}{\textbf{DEAP Dataset}} \\ \hline
Shu et al.\cite{RN630} & 2017 & \makecell[c]{\emph{Restricted Boltzmann Machine (RESPIRATION BELTM) + SVM} \\ \emph{EEG(32), EOG, EMG, GSR, Respiration belt, Temperature, Plethysmograph}} & 2 & 62.65 \\ \hline
Yin et al.\cite{RN312} & 2017 & \makecell[c]{ \emph{Multiple-fusion-layer based Ensemble classfier of Stacked Autoencoder (MESAE)} \\ \emph{EEG(32), EOG, EMG, GSR, Respiration belt, Temperature, Plethysmograph}} & 2 & 83.61 \\ \hline
Kwon et al.\cite{RN520} & 2018 & \makecell[c]{ Fusion CNN \\ EEG(32), GSR} & 4 & 73.43 \\ \hline
Hassan et al.\cite{RN425} & 2019 & \makecell[c]{ DBN+Fine Gaussian Support Vector Machine (FGSVM) \\ EMG, GSR, Plethysmograph} & 5 & 84.44 \\ \hline
Ma et al.\cite{RN678} & 2019 & \makecell[c]{ \emph{Multimodal Residual LSTM} \\ \emph{EEG(32), EOG, EMG, GSR, Respiration belt, Temperature, Plethysmograph}} & 2 & 92.59 \\ \hline
Huang et al.\cite{RN679} & 2019 & \makecell[c]{ Ensemble CNN \\ EEG(32), EOG, GSR, Respiration belt} & 4 & 82.92 \\ \hline
Zhou et al.\cite{RN627} & 2019 & \makecell[c]{ Convolutional Auto-Encoder (CAE) + FCNN \\ EEG(14), GSR, Temperature, Respiration belt} & 4 & 92.07 \\ \hline
Bagherzadeh et al.\cite{RN629} & 2019 & \makecell[c]{ \emph{Parallel Stacked Autoencoders (PSAE)} \\ \emph{EEG(32), EOG, EMG, GSR, Respiration belt, Temperature, Plethysmograph}} & 4 & 93.60 \\ \hline
\textbf{Our Model} & \textbf{} & \makecell[c]{ \textbf{Multiscale CNNs + Biologically Inspired Decision Fusion Model} \\ \textbf{EEG(32), EOG, EMG, GSR, Respiration belt, Temperature, Plethysmograph}} & \textbf{4} & \textbf{98.52 $\pm$ 0.09} \\ \hline
\multicolumn{5}{|c|}{\textbf{AMIGOS Dataset}} \\ \hline
Yang et al.\cite{RN683} &2019 & \makecell[c]{ \emph{Attribute-invariance loss embedded Variational Autoencoder (AI-VAE)} \\ \emph{EEG(14), ECG, GSR}} & 2 & 64.65 \\ \hline
Granados et al.\cite{RN486} & 2019 & \makecell[c]{ 1D CNN + FCNN \\ ECG, GSR} & 4 & 65.25 \\ \hline
Shukla et al.\cite{RN682} & 2019 & \makecell[c]{ Hand-crafted Features + SVM \\ GSR} & 2 & 84.83 \\ \hline
Harper et al.\cite{RN681} & 2019 & \makecell[c]{ Bayesian Deep Learning Framework \\ ECG} & 2 & 90 \\ \hline
Li et al.\cite{RN677} & 2020 & \makecell[c]{ \emph{Attention-based Bidirectional Long LSTM-RNNs + DNN} \\ \emph{EEG(14), ECG, GSR}} & 2 & 81.35 \\ \hline
\textbf{Our Model} & \textbf{} & \makecell[c]{ \textbf{Multiscale CNNs + Biologically Inspired Decision Fusion Model} \\ \textbf{EEG(14), ECG, GSR}} & \textbf{4} & \textbf{ 99.89 $\pm$ 0.05} \\ \hline
\end{tabular}
\begin{tablenotes}
        \footnotesize
        \item The EEG (32) means 32 channels EEG signals be used.
        \item Italics indicate that the signals used in the previous studies are consistent with ours.
	\end{tablenotes}
\end{threeparttable}
\end{table*}

\section{Conclusions}
In this paper, we have proposed a Multiscale Convolutional Neural Network Model (Multiscale CNNs) and a Biologically Inspired Decision Fusion Model for affective states recognition from multimodal physiological signals. The High Scale CNN and Low Scale CNN in the Multiscale CNNs are utilized to predict affective states based on EEG signals and peripheral physiological signals respectively. Then the biologically inspired decision fusion model integrates the results from Multiscale CNNs to get the final result. The fusion model improves the accuracy of affective states recognition significantly compared with the result on single-modality signals, and it could be applied to other similar problems. We also compared the performance of the decision fusion model and the plurality voting method which widely used in other decision fusion studies. Compared with it, the biologically inspired decision fusion model considers the correlation between various class labels. Then the reliability of each single-modality signals is calculated by this correlation and the classification probability from physiological signals. Finally, the decision is made by the more reliable modality information while it retains other modalities information. In addition, the results demonstrate that the Multiscale CNNs are effective for affective states recognition using single-modality signals, especially the EEG-based High Scale CNN affective states recognition network. The primary fusion results based on four frequency bands EEG signals proves that the original EEG signals contain enough information and could be used directly for affective states classification. The primary fusion result of peripheral physiological signals shows that a robust classification of human affective states using peripheral physiological signals without EEG signals is possible. 

The proposed method could predict affective states with high accuracy in the public database DEAP and AMIGOS. Although this method can not predict the complex construct of emotions\cite{6797872,kleinginna1981a,scherer2000psychological}, it lays a foundation for real-world applications based on physiological-based affective states recognition. With the help of wearable devices and edge computing, it is expected to be applied in the real-world, such as monitoring negative affective states, identifying autism spectrum disorders, etc. Compared with other signals such as facial expression, gestures or speech, the physiological signals are not easy to be controlled subjectively and camouflaged. And the physiological-based affective states recognition method pays more attention to privacy and the application scenario is more flexible. This method does not need to use camera to collect people's facial expressions or gesture, or use a tape-recorder to record people's voice, which can minimize the impact on people's privacy. Besides, this method does not need to deploy cameras or tape-recorders in the scene in advance, nor does it need to make people always in the recording angle range of cameras or tape-recorders. People only need to wear wearable devices that can collect physiological signals, such as Wearable EEG Sensor, ECG Sensor, Skin Temperature Sensor and Blood Pressure Sensor, etc., so they can move freely and not being limited to specific scenes. Finally, compared with facial expression, gesture or speech, physiological signal data has lower dimensions and requires less computational resources. There are still many challenges in this method. In the future, we will reduce the computing resources by optimizing the network structure, and design more reliable and comfortable wearable physiological signal acquisition devices to promote the application of this method in the real-world.

\bibliography{scibib}
\bibliographystyle{unsrt}
\end{document}